\documentclass{aa}
\usepackage{graphicx}
\begin{document}
\authorrunning{Laycock et al}
\titlerunning{New Transient SMC Pulsar}

   \title{Discovery of a new Transient X-Ray Pulsar in the Small Magellanic Cloud}

   \author{S. Laycock \inst{1}, R. H. D. Corbet\inst{2,3}, D. Perrodin\inst{2,3}
   , M. J. Coe \inst{1}, F. E. Marshall \inst{2}, C. Markwardt \inst{2,4}
   }

   \institute{Dept. of Physics and Astronomy, University of Southampton,
              Highfield, Southampton, SO17 1BJ, UK \\
              \email{sgtl@astro.soton.ac.uk}
         \and
             NASA Goddard Space Flight Center, Greenbelt, MD 20771, USA\\            
         \and
         	Universities Space Research Association \\
	\and
		Department of Astronomy, University of Maryland, College Park, MD 20742                 
             }

   \date{Received August 23, 2001; accepted January 28, 2002}

   \abstract{  
Systematic {\it Rossi X-Ray Timing Explorer (RXTE)} observations of the Small Magellanic Cloud (SMC) 
have revealed a previously unknown transient X-ray pulsar with a pulse period of 95s.  
The 95s pulsar, provisionally designated XTE SMC95, was detected in three Proportional Counter Array (PCA)
observations during an outburst spanning 4 weeks in March/April 1999. The pulse profile is double peaked 
reaching a pulse fraction of $\approx$0.8. The X-ray spectrum is well represented by an absorbed power-law with a 
photon index of 1.4 and mean unabsorbed flux of $\ga$ 8.95$\times 10^{-11}$ ergs cm$^{-2}$ s$^{-1}$ (3 - 25 keV).
The source is proposed as a Be/neutron star system on the basis of its pulsations, 
transient nature and characteristically hard X-ray spectrum. The 2-10 keV X-ray luminosity implied by 
our observations is $\ga$ 2$\times$10$^{37}$ergs s$^{-1}$ which is consistent with that of normal
outbursts seen in Galactic systems. This discovery adds to the emerging picture of the SMC
as containing an extremely dense population of transient high mass X-ray binaries.   
   }
\maketitle

%

\section{Introduction}

Observations of the SMC by {\it RXTE, ASCA} \& {\it SAX} have detected an increasing number of new transient X-ray 
pulsars over the last few years, e.g. Haberl \& Sasaki (\cite{hs00}), Yokogawa et al (\cite{Yokogawa00}). With no sign of a decline in 
this trend, the picture emerging 
from this flurry of new identifications and rediscoveries of sources seen by {\it Einstein} and {\it ROSAT} is 
that the SMC contains a large population of X-ray binary systems, of which only a small fraction are 
active at any one time. The SMC provides the closest available approximation to a luminosity-limited
sample of X-ray pulsars due to its low extinction and small size. The depth of the SMC is small relative to its 
distance from us, placing all of the pulsars at effectively the same distance.  {\it RXTE}'s proportional counter 
array is well suited to studying this population since its field of view, sensitivity and timing resolution 
allow simultaneous monitoring of all sources in a significant fraction of the SMC.     

Observational data from across the spectrum has provided a coherent 
classification of X-ray binaries, both in the Galaxy and Magellanic clouds. The principal 
distinction is between high and low mass stellar counterparts, forming a distinction that 
is also generally coincident with that between sources exhibiting
coherent X-ray pulsations, and those that do not. In high mass X-ray binary (HMXB) 
systems the optical emission is dominated by the star, while in the latter low mass (LMXB) case, 
emission from an accretion disk around the compact object dominates. The HMXBs fall into
only two subclasses on the basis of their optical classification, those with an OB supergiant (SG)
optical counterpart and those involving a Be emission-line star, which account for the majority
of HMXBs. These two subgroups broadly coincide with two distinct modes of X-ray behaviour, 
persistent and transient sources. 

Systems containing a supergiant are persistent sources of 
X-rays which are usually modulated by eclipses due to the short orbital period and correspondingly
small size of the system. In very short period systems the primary overfills its Roche-lobe and the overflowing
matter is forced by its high angular momentum into a stable accretion disk around the neutron star. 
In more widely separated systems, persistent emission arises by accretion from a roughly spherically 
symmetric stellar wind. Pulse-timing measurements 
suggest that long-term stable accretion disks do not form in these systems, although
short-lived torque-reversals observed in many systems (Bildsten et al \cite{Bildsten97}) imply the  
formation of transient accretion disks.    
  
Transient X-ray sources account for the majority of HMXBs. The neutron star's (NS) optical companion is 
generally a Be III-V star exhibiting a strong infrared flux excess. Long term monitoring of these 
systems has resulted in a `standard model' in which the Be star, rotating close to its
breakup speed, is surrounded by a dense equatorial disk of cool material. This disk is prone 
to successive phases of building and ejection on timescales of 1-3 years (Coe \cite{Coe00a}). 
The neutron star, in a wide (few 10s - 100s of solar radii) eccentric orbit around the Be star, 
periodically intercepts the ejected material
leading to the formation of an accretion disk and X-ray emission. This general model explains 
the regular X-ray outbursts of Be/NS systems, which are modulated at the orbital period and
normally occur close to periastron passage of the neutron star. The situation is complicated
by changes in the circumstellar disk which alter the density of material being encountered by 
the neutron star. This leads to long epochs (several orbital periods) of heightened X-ray activity
often accompanied by strong torque reversals and prolonged periods of spin-up. Over the long-term,  
such activity leads ultimately to a dynamic equilibrium between the orbital and neutron star spin periods
since the pulsar experiences negative as well as positive torques.
The equilibrium is manifested observationally in the P$_{Pulse}$ vs P$_{Orbit}$ diagram (Corbet et al \cite{Corbet99}). 
See also Corbet (\cite{Corbet86}), Stella et al (\cite{Stella86}) and Bildsten et al (\cite{Bildsten97}) for
detailed explanations. 

With the addition of recent discoveries there are now 25 X-ray pulsars known  
in the SMC. Of these X-ray sources, the majority have been identified as belonging 
to Be systems rather than supergiant systems (Haberl \& Sasaki \cite{hs00}). 
Comparing these figures to the population in the Galaxy in which about 70 HMXBs are known, 31 
identified with Be optical companions, the SMC emerges as containing a particularly dense population 
of Be X-ray binaries. Furthermore of the 14 supernova remnants (SNR) identified in the SMC
none is Crab-like (Yokogawa et al \cite{Yokogawa00}). Since Crab-like SNR are believed to form from the explosion of a single 
massive star it is likely that the evolutionary history of the SMC has resulted in a large
ratio of massive binary stars to isolated massive stars. 
When the comparison is scaled by relative masses of the SMC and Galaxy it is clear that the population 
density of HMXBs in the SMC is higher even than the Galaxy's most populous regions. Many authors 
e.g. Stavely-Smith et al (\cite{Stavelysmith97}) and Popov et al (\cite{Popov98}) 
have proposed an evolutionary scenario for the SMC in which an extremely active epoch of star
formation occurred relatively recently (within 5 Myr), leading to the large proportion of young, massive stars.
In support of this theory are 4 key observations: (1) Large number of supernova remnants approximately 5 Myr old; 
(2) The large number of HMXBs; (3) Apparent absence of low mass systems which would represent an 
older stellar population. (These still remain elusive despite the current generation of satellites' ability 
to detect them if they are similar to galactic LMXBs); (4) The ratio of Be/B type stars in the SMC is 0.39 
(Maeder et al \cite{mgm99}) compared to 0.16 in the Galaxy, implying a younger population.


\section{The RXTE Observations}

The {\it RXTE} Proportional Counter Array (PCA) (Jahoda et al \cite{Jahoda96}) consists of 5 collimated, 
Xenon filled, proportional counter units (PCU), each with a collecting area of 1300 cm$^{2}$. The field of view 
of the PCA is approximately 2\degr FWZI with a triangular collimator response as a function of 
off-axis viewing angle. The PCA has 3 detector-anode layers, this arrangement greatly increases the signal/noise by 
physically separating the detection of low, medium and high energy X-rays. In certain data modes (Good Xenon and 
Standard 2) this separation is maintained in the telemetry, allowing data from each layer and PCU to be analysed 
independently. The highest timing and spectral resolution are obtained in Good Xenon mode, in which the energy and 
arrival time of of every detector event is registered with 1$\mu$s timing accuracy and 256 channel energy resolution. 
Standard 2 mode provides time-binned data with 16s resolution and 129 energy channels. Both of these data modes were 
used in our analysis, as explained in Section 3. 
 
\begin{figure}
   \centering
   \includegraphics[width=7.5cm]{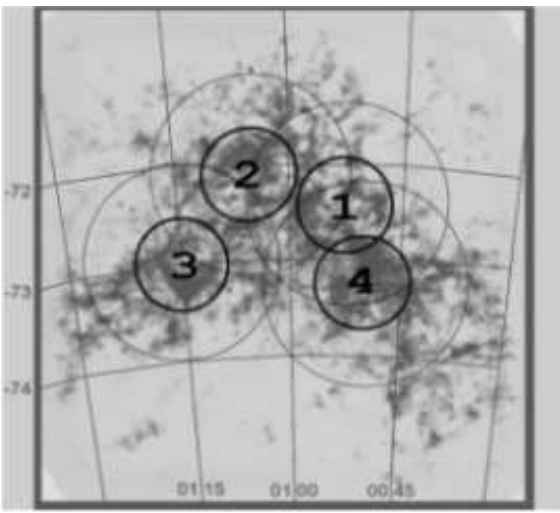}
      \caption{fields of view used in our systematic monitoring program, superimposed
        on an H1 radio map of the SMC (Stanimirovic et al \protect\cite{Stanimirovic99}). The
	precise pointing centres are given in Table~\ref{obs}. The thick and thin 
	circles represent the half-power width and FWZI of the PCA respectively. 
	Data from the 4$^{th}$ position are not presented in this work.}
   \label{smcpic}
   \end{figure}

{\it RXTE} observations were obtained in the course of our regular pulsar search program 
of the SMC. Approximately every week, 3 ksec observations were made centred on  one of 4 partially
overlapping positions (See Table~\ref{obs} and Figure~\ref{smcpic}) in the `bar' and `wing' regions 
of the SMC where young objects are expected to be found (Kahabka \& Pietsch 1996) This program began in January 1999 and has 
allowed detailed monitoring of the long-term behaviour of pulsars in the vicinity of SMC X-3 (Lochner et al \cite{Lochner96}). 
At the time of the observations presented here, all four pointing positions were being routinely observed, 
with about three quarters of the time concentrated on the region around SMC X-3 (position 1). However 
since early 2000 our strategy has changed to allow us to observe the single position centred on SMC X-3 
with much greater regularity, since most of the pulsars seen during this program and by 
{\it ASCA} (Haberl \& Sasaki \cite{hs00}) have been visible in this position.

Two observations of position 1 taken seven days apart during March 1999 revealed the existence of a transient 
X-ray pulsar (XTE SMC95) whose 95s pulse period and pulse profile we were able to determine. The next position 1
observation two weeks later revealed weak pulsations, after which the source faded from view.  

Although individual pointed detections of a transient source are unable to constrain its position, 
the only impact of this positional uncertainty on PCA data analysis is uncertainty in absolute flux 
determination. Given the high timing resolution and S/N typically obtainable for SMC pulsars, this 
limitation is unimportant in the identification of new sources and the study of pulse profiles and 
pulse-frequency evolution, which are the primary aims of this program.


\section{Data Reduction}
 
Good Xenon mode data from the PCA were used to 
construct lightcurves and spectra. 4 PCUs were active during the observations presented here.
In order to maximise signal-to-noise ratio, only events from the top anode layer of each active proportional counter 
unit (PCU) were included and the analysis was restricted to the 3-25 keV energy range. The PCA is sensitive to 
X-rays between 2-60 keV, although for faint sources its spectral response seems unreliable below 3 keV.
The X-ray flux from pulsars falls off rapidly above 10 keV.  

Additional filtering criteria were applied to exclude data collected during periods of high background occurring shortly 
after SAA passage and short data segments coinciding with start and finish of pointing slews. Background  
subtraction was performed using the L7 PCA faint background models (gain epochs 3 and 4) provided by 
the {\it RXTE} Guest Observer Facility. These models use the Standard 2 mode PCA data to determine various
parameters used in estimating the background count-rates and spectra in the specific PCU and anode-layer configuration being 
used for each observation. Background corrected lightcurves were then constructed at 100 ms timing resolution, with bin times 
corrected to the solar-system barycentre and the count-rates normalised to counts PCU$^{-1}$s$^{-1}$.  A detailed
description of the L7 background model is given by Edelson \& Nandra (\cite{EN99}). 


\section{Timing Analysis}
In order to search for pulsars in the datasets, we calculated power spectra covering a period range 
0.1-1000 seconds at a resolution of 0.1 mHz.
Coherent pulsations were initially detected in observation 3 and power spectra showing the 
positive detection of XTE SMC95 are shown in Figures~\ref{powspec1} \&~\ref{powspec2}.
These power spectra show several harmonics of the fundamental 
spin frequency, whose relative strengths changed between the two observations. The fundamental and 
harmonic frequencies were weighted by their spectral power to obtain a best fit to the true period. The
uncertainty in the period was calculated by propagating the error (FFT resolution) from each
harmonic. Two harmonics were used to establish the period in observation 3, Three in observation 4 (see
Figures~\ref{powspec1} \&~\ref{powspec2}).

After determining the pulse period and associated observational errors, average pulse profiles were 
created for each of the observations. The profiles are shown in Figure~\ref{profiles95}, while key observational 
parameters derived from them are given in Table~\ref{summary}. The mean fluxes have formal uncertainties
of 0.03 counts PCU$^{-1}$s$^{-1}$ before taking into account contribution from interfering sources and uncertainties 
in the background model. Using observations for our monitoring program in which no pulsations were detected, an estimate
was made of the contribution from unpulsed X-ray sources in the field, such as supernova remnants. The mean background
corrected 3-10 keV flux averaged over it 3 lowest values for observations of Position 1 is 0.98$\pm$0.05 Count
PCU$^{-1}$s$^{-1}$ with an excess variance in the count rates of 13\% over the expected Poisson level. Thus the true 3-10
keV pulse fraction in observation 3 (See Figure~\ref{profiles95}) was $\approx$0.8.  

A weak 95s pulsation was detected in observation 6, however the flux was dominated by 91s pulsations from the nearby 
pulsar AX J0051-722. Due to the closeness of the two pulse periods, direct measurement of the pulse amplitude of SMC95 was
not possible. Instead simulated data sets were created using the pulse profile obtained in observation 4 as a template and
varying its pulse amplitude. These simulated data were used to estimate the pulse amplitude required to recreate the 95s
peak seen in the observation 6 PDS. The resulting upper limit (See Table~\ref{summary}) is approximately one third
the amplitude seen in obs.4 and is found to
be nearly 10 times the amplitude required if the modulation were sinusoidal, a possibility supported by the lack of
harmonics in the PDS. Considering the mean count-rate in the observations, it is clear that most of the flux in observation
6 was due to AXJ0051-722. Thus the brightness of SMC95 is constrained to be in the lower end of the range determined from
the timing alone.

\section{Spectrum analysis}
\begin{figure}
\centering
\includegraphics[angle=-90,width=7.5cm]{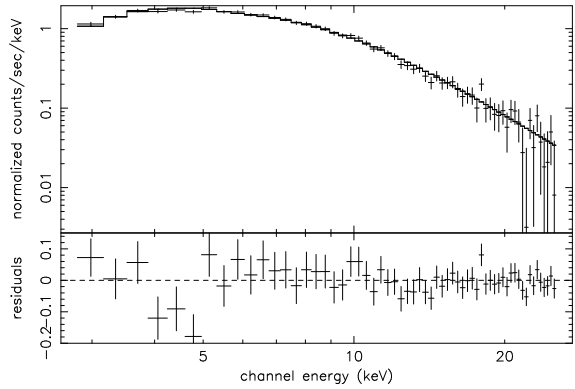}
\caption{{\it RXTE} PCA spectrum obtained during observation 3, when XTE SMC95 was the 
only detectable pulsar in the field of view. Solid line is an absorbed power-law fit, residuals are
shown in the lower panel. See Table~\ref{spec} for details of the spectral parameters.}
\label{spec08}
\end{figure}

PCA spectra were obtained from the first two observations containing XTE SMC95, the spectra are 
well represented by an absorbed power-law typical of Be/X-ray binary pulsars (e.g. White, Swank \& Holt \cite{White83}). 
The fitted model parameters are given in Table~\ref{spec} for the two observations and are in agreement
to within errors. Figure~\ref{spec08} shows the spectrum from observation 4, along with the fitted model and residuals. 
Uncertainties in the fitted parameters represent the limits of the 90 \% confidence interval. 
The unabsorbed source fluxes were obtained from the fitted spectral models. Uncertainties in the absolute flux
are dominated by the unknown source position and hence collimator response. Since both observations were made with the 
same pointing, the uncertainty is systematic.

\section{Source Location}
PCA slew data were used to constrain the source position.  The PCA
detectors are often on and accumulating data as the spacecraft is
manoeuvered on-source during the start of an observation.  As the
source moves through the collimator field of view, it samples
different values of the collimator response function, which modulates
the detected count rate.  Given a trial position and intensity for the
source, a model light curve can be constructed using a tabulated
response function, and compared to the observed light curve.  The
position and intensity are then varied iteratively using least squares
fitting until the best-fit parameters are obtained.  Ideally, two
scans completely over the source are needed, in orthogonal directions,
to constrain the source position with the highest precision.  When, as
in this case, the scans are not complete, and not at right angles, the
error box will be larger and asymmetric.

Light curves for XTE SMC95 were extracted from the 16-second Standard2
data in the 2--10 keV band (PCA top layer only) for observations 3 and
4.  The light curves included approximately 200 seconds of slewing and
on-source time each.  The model included a term for the PCA
background, which was fixed at the value determined from the
PCABACKEST model for faint sources.  The intrinsic variability of the
pulsar can be confused with variations of source position.  To
compensate for this, the light curve error bars were increased
proportionately by 30\% (this number was also scaled according to the
collimator response at each time sample).  The value of 30\% was
determined by fitting a constant intensity during the pointed portion
of the observation, and forcing $\chi^2_\nu = 1$, and represents the
effective fractional r.m.s. variation of the pulsar on 16 second
timescales.

The best fitting position was $\alpha = 13.36^\circ$, $\delta =
-72.821^\circ$ (J2000), with a reduced chi-square value of 1.02.  The position
uncertainty is a thin strip shown in Figure~\ref{errorbox}.  Confidence contours
were estimated by taking $\Delta\chi^2 = {6.2, 9.2, 11.8, 18.4}$,
which formally represent confidence levels of $2\sigma$, 99\%,
$3\sigma$, and 99.99\% respectively, for two parameters.  However,
because the variability of the pulsar is non-statistical and
non-trivial to model, we favor assigning the $3\sigma$ contour as the
smallest reasonable error box.  The semi-minor and semi-major axes of
the error box are 5 arcmin and 41 arcmin respectively.  Since the
source was not detected in PCA observations at pointing positions
number 2 and 4, which are east and south of pointing position number
1, respectively, it is more likely that the source is located to the
northwest of the best-fit position.

The true intensity of the source depends on its position within the
PCA collimator response function.  At the best-fit position, the
collimator response is $\sim 60$\%, implying that the source is $\sim
65\%$ brighter than implied by the counting rate observed by the PCA
at pointing position number 1.  Given the full error box described
above, the collimator response could potentially be in the range of
10\%--75\%.  If the source is northwest of the best-fit position, the
collimator response is in a more restricted range of 60\%--75\%, and
then it is likely that the observed rate is close to the actual source flux
, within a factor of $\sim 2$.

\section{Duration of the Outburst}  
As Table~\ref{obs} shows, XTE SMC95 was not detected in observations made before and after those presented here.
The closest before and after observations were made at pointing positions 2 \& 3 respectively, one week either side of 
the detected outburst. As Figure~\ref{smcpic} shows, these observations provide no useful information due to their 
limited overlap, other than possibly constraining XTE SMC95 to lie towards the west in the position 1 field of view. 
The closest position 1 observations occurred within two weeks of the outburst.
Analysis of power spectra obtained in these bracketing observations allow us 
to place upper limits on the X-ray flux during apparent quiescence. From this analysis we find that XTE SMC95
pulse amplitude was below our approximate 3$\sigma$ detection threshold of 0.25 counts PCU$^{-1}$s$^{-1}$ in 
observations 1 and 7, where this figure is directly comparable with the measured fluxes in observations 3 \& 4.


\section{Discussion}

Most galactic Be/X-ray binaries increase in luminosity
by a factor of $\ga$100 during normal outburst (Negueruela \cite{Negueruela98}) reaching luminosities around 
10$^{37}$erg s$^{-1}$. Taking the distance to the SMC as 63 kpc (Groenewegen et al \cite{Groenewegen00}), 
the fluxes given in Table~\ref{spec} imply a 2-10 keV luminosity for XTE SMC95 of 
at least 2$\times$10$^{37}$ erg s$^{-1}$. Thus during this outburst the X-ray luminosity of XTE SMC95 was comparable to 
that of Galactic Be/NS systems.    
The photon indices obtained from our spectral fits are slightly steeper than those normally seen in Galactic Be/NS HMXBs, 
values $\la$ 1.0 being typical in the 10-20 keV range (White et al \cite{White83}). Softer (steeper) spectra have been 
associated with less luminous sources based on studies in the Galaxy (White et al \cite{White83}). 
Comparing our results (Table~\ref{spec}) to the recent outburst of SMC X-2 which was both more luminous and 
had a harder spectrum: luminosity $\approx$10$^{38}$ergs s$^{-1}$, spectral index 0.7-1.0 (Corbet et al \cite{Corbet01}) this trend 
looks likely to hold in the SMC.  

\begin{figure}
\centering
\includegraphics[angle=-90,width=7.5cm]{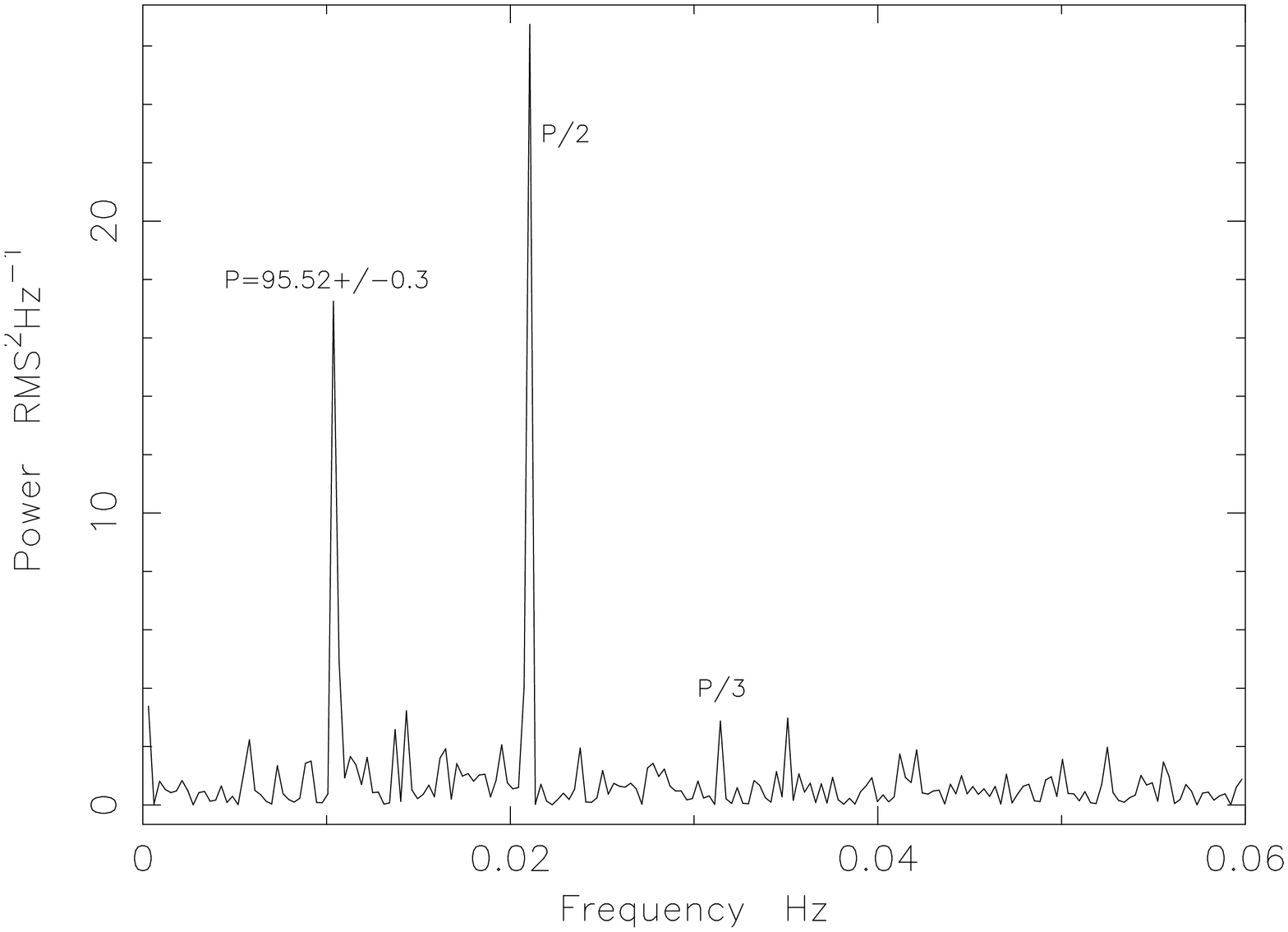}
\caption{PCA 3-25 keV power spectrum showing fundamental pulse frequency and harmonics 
	of XTE SMC95 in observation 3.}
\label{powspec1}
\end{figure}

\begin{figure}
\centering
\includegraphics[angle=-90,width=7.5cm]{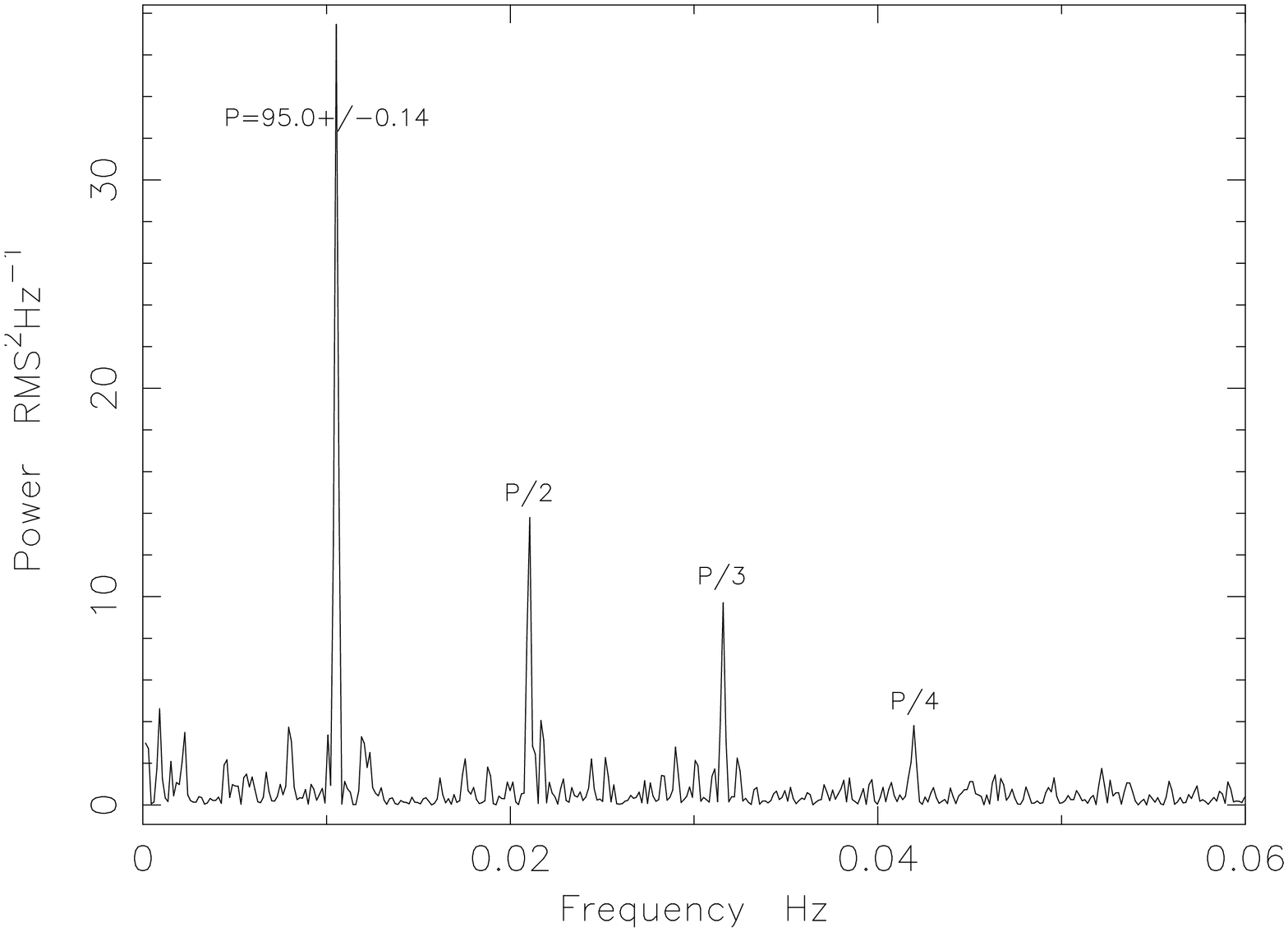}
\caption{PCA 3-25 keV power spectrum showing fundamental pulse frequency and 
	multiple harmonics of XTE SMC95 in observation 4.}
\label{powspec2}
\end{figure}

\begin{figure*}
\centering
\includegraphics[width=16cm]{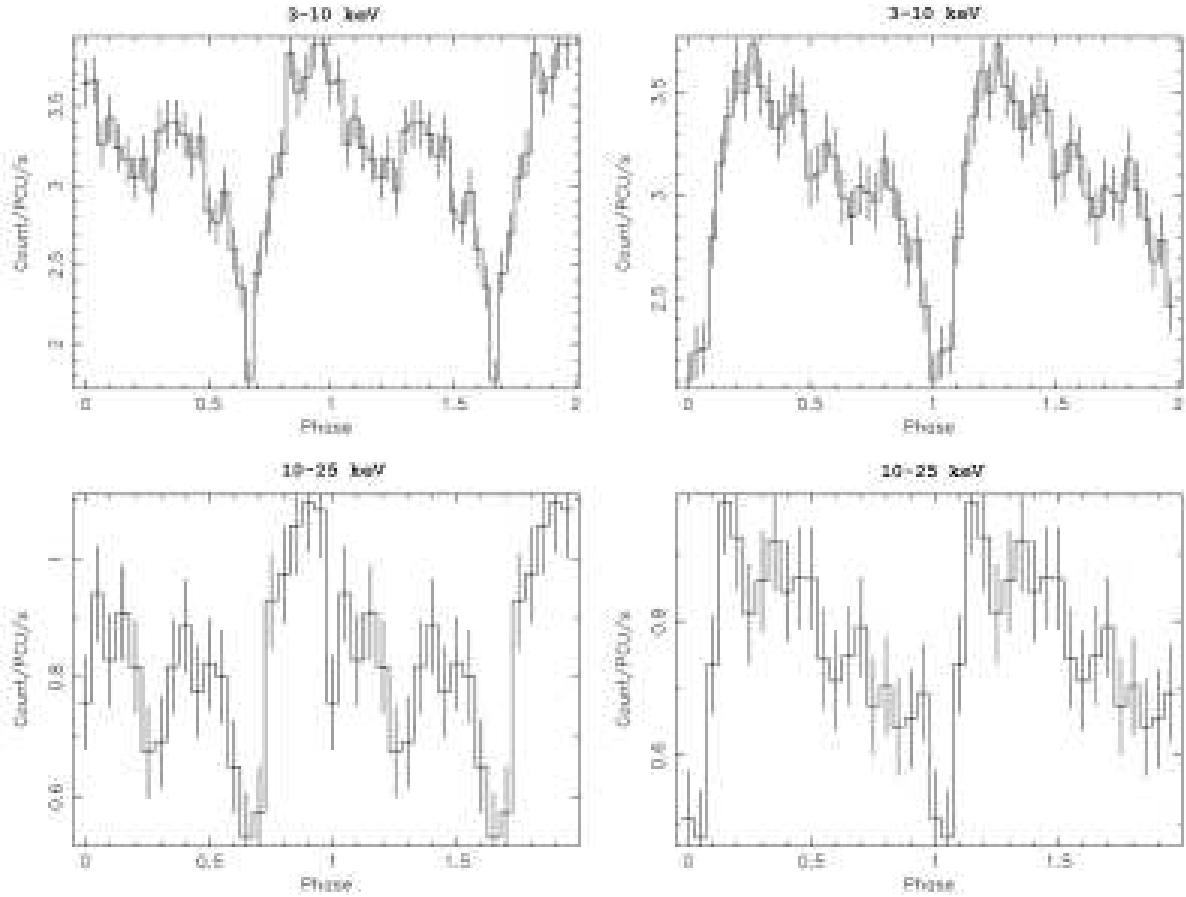}
\caption{Pulse profiles for XTE SMC95 in 3-10keV and 10-25keV energy bands 
	(Arbitrary phase T$_{0}$=MJD51248). 
	Left column: observation 3. Right column: observation 4.}
\label{profiles95}
\end{figure*}

XTE SMC95 exhibited varying structure in its pulse profile. Pulse profiles 
obtained in two separate energy bands 3-10 keV and 10-20 keV shown in Figure~\ref{profiles95}
indicate the pulse-shape changed between the two observations 
(3 \& 4), with the second pulse peak weakening relative to the first, accompanied by a decrease in pulsed flux. The
significant shift in relative strength of the first and second power spectrum harmonics between the two observations 
(see Figures~\ref{powspec1} \&~\ref{powspec2}) confirms that the profile change is real 
and not an artifact of an incorrect profile-folding period. 

A small decrease in the pulse period was detected between the two observations
of XTE SMC95 one week apart, implying a pulse period derivative (8.6 $\pm$ 5.5) $\times$10$^{-7}$ s s$^{-1}$. 
This change is attributed to an unknown combination of spin up and orbital motion depending on the inclination of the system. 

For Be/NS systems with P$_{pulse}$ around 95s, the Corbet diagram (Corbet \cite{Corbet86}) 
suggests likely orbital periods in the 50-100d range. XTE SMC95 was detected for three weeks, only a small fraction of its 
likely orbital period. Whether this outburst was related to orbital motion or long-term changes (disk-loss events)
in a circumstellar disk may be determined by continued monitoring.

\begin{figure}
\centering
\includegraphics[angle=-90,width=7.5cm]{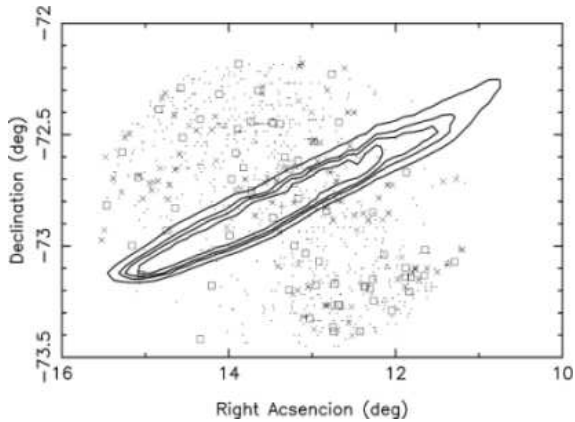}
\caption{Confidence contours for the position of XTE SMC95.
Formal confidence levels are $2\sigma$, 99\%, $3\sigma$, and 99.99\%
(see Section 6).  The best fit position is shown as a ``+''
symbol, and the position of the PCA at pointing position number 1 is
indicated with a circle. ROSAT sources are indicated by squares, luminous O/B
stars crosses, H$\alpha$ stars dots. See Section 8}
\label{errorbox}
\end{figure}

A search of the ROSAT source catalogue (Haberl et al \cite{hfp00}) reveals 4 sources within the 3$\sigma$ confidence region
(See Figure~\ref{errorbox}). Of these sources, RX J0053.8-7252 is classified as a candidate X-ray
binary and RX J0052.6-7247 a background AGN the other sources are a
foreground star and a Supersoft X-ray source. A similar search for known Be stars was made in the
same region. Searching the AzV (Azzopardi \& Vigneau \cite{azv82}) \& LIN (Lindsay \cite{lin}) catalogues for
spectral types: O8-B2, I-V, gave 7 stars within the 3$\sigma$ contour. This range of
spectral types was chosen to match
the known distribution for optical counterparts to HMXBs (Negueruela \cite{Negueruela98}). The AzV
catalogue is 80\% complete to limiting magnitude B=14.3, considering the distance to the SMC the sample
is strongly biased toward early-type giants. Thus main sequence Be stars (Luminosity class IV-V) are likely 
to be under represented in the sample. The catalogue of H$\alpha$ emission-line objects of Meyssonnier \&
Azzopardi (\cite{ma93}) lists 69 stars within the 3$\sigma$ contour. From the argument given above, the
majority of these are likely to be O/Be stars.

\begin{acknowledgements}
We thank the referee, W. Heindl for constructive comments.
This research has made use of the SIMBAD database, operated at CDS, Strasbourg, France. 
SL is supported by a PPARC studentship.
\end{acknowledgements}

\begin{table}
 \caption{Dates and Positions of Observations}
 \label{obs}
 \begin{tabular}{lcccc}
 Obs.   Date        & Position  &  Duration /ks & Source        \\
 1      25/2/99     &  1        & 2.20          &  -            \\
 2       4/3/99     &  2        & 3.38          &  -            \\
 3      11/3/99     &  1        & 3.13          &  XTE SMC95    \\
 4      18/3/99     &  1        & 3.49          &  XTE SMC95    \\
 5      26/3/99     &  3        & 3.76          &  -            \\
 6       1/4/99     &  1        & 3.78          &  XTE SMC95    \\
 7       9/4/99     &  1        & 4.49          &  -            \\
 8      22/4/99     &  1        & 2.98          &  -            \\
\end{tabular}

\medskip
 {\em Position 1:} RA=13\fdg471; Dec=-72\fdg445, {\em Position 2:} RA=16\fdg25; Dec=-72\fdg106,
 {\em Position 3:} RA=18\fdg75; Dec=-73\fdg091, {\em Position 4:} RA=12\fdg686; Dec=-73\fdg268 
 (Equinox 2000.0)  
\end{table} 


\begin{table}
 \caption{Summary of Timing Results}         
 \label{summary}
 \begin{tabular}{lccc}
 Obs.        & Pulse period      &  Pulse amplitude$^{\dag}$ & Mean flux$^{\dag}$    \\
 number      & seconds           &  Ct PCU$^{-1}$s$^{-1}$    & Ct PCU$^{-1}$s$^{-1}$ \\ 
   3         &  95.49$\pm$0.05   &  2.72$\pm$0.26            &  4.34                 \\
   4         &  95.03$\pm$0.03   &  2.27$\pm$0.25            &  4.08                 \\
   6         &  95.0$\pm$0.1     &    0.2-1.0$^{\ast}$       &  3.79               \\
\end{tabular}

\medskip
\dag Background corrected 3-25 keV. $\ast$ See Section 4 
\end{table} 

\begin{table}
 \caption{Spectral fits to observations}          
 \label{spec}
 \begin{tabular}{lllccc}
Obs. & $N_{H}$           &  $\alpha$         & flux$^{\dag}$ & L$_{X}^{\ddagger}$      & $\chi_{\nu}^{2}$  \\
     & $\times10^{22}$   &                   &               &                         &             \\
3    & 1.82 $\pm$ 0.89   & 1.41 $\pm$ 0.23   & 8.95          & 2.14                    & 0.86        \\
4    & 2.66 $\pm$ 0.89   & 1.50 $\pm$ 0.07   & 8.43          & 2.15                    & 0.89        \\
\end{tabular}

\medskip
\dag total unabsorbed model flux $\times$10$^{-11}$erg cm$^{-2}$s$^{-1}$ (3-25 keV) \\
$\ddagger$ 2-10 keV model luminosity $\times$10$^{37}$erg s$^{-1}$ at 63 kpc \\
N.B. These are upper limits with a factor $\sim$5 uncertainty in flux \& L$_{X}$ due to unknown collimator response. (No
corrections applied) \\
$\chi_{\nu}^{2}$ for 58 degrees of freedom.
\end{table} 

%

\end{document}